\documentclass[
twocolumn,
floatfix,
longbibliography,
superscriptaddress,
amssymb,
amsmath
]{revtex4-2}

\usepackage[dvips]{graphicx}% Include figure files
\usepackage{epsf}
\usepackage{dcolumn}% Align table columns on decimal point
\usepackage{bm}% bold math
\usepackage{bm, color}
\usepackage{bbm}
\usepackage{lineno}
\usepackage{braket}
\usepackage{amsbsy}
\usepackage{xr}
\usepackage{siunitx} 
\usepackage{easyReview}
\usepackage[%dvipdfmx,
colorlinks=true,
linkcolor=magenta,
urlcolor=magenta,
citecolor=magenta,
filecolor=blue,
pagecolor=blue,
linktocpage=true,
bookmarks=true,
bookmarksnumbered=true
]{hyperref}

\begin{document}

\title{$d$-Wave Flat Fermi Surface in Altermagnets Enables Maximum Charge-to-Spin Conversion}
%\add{(1) $d$-Wave Flat Fermi Surface in Altermagnets Enables Record Charge-to-Spin Conversion\\
%	(2) Maximized Charge-to-Spin Conversion via $d$-Wave Flat Fermi Surface in Altermagnets\\
%	(3) $d$-Wave Flat Fermi Surface in Altermagnets: A Novel Route to Maximize Charge-to-Spin Conversion
%}

\author{Junwen Lai}
\affiliation{Shenyang National Laboratory for Materials Science, Institute of Metal Research,Chinese Academy of Sciences, Shenyang 110016, China.}
\affiliation{School of Materials Science and Engineering, University of Science and Technology of China, Shenyang 110016, China.}

\author{Tianye Yu}
\affiliation{Shenyang National Laboratory for Materials Science, Institute of Metal Research,Chinese Academy of Sciences, Shenyang 110016, China.}

\author{Peitao Liu}
\affiliation{Shenyang National Laboratory for Materials Science, Institute of Metal Research,Chinese Academy of Sciences, Shenyang 110016, China.}
\affiliation{School of Materials Science and Engineering, University of Science and Technology of China, Shenyang 110016, China.}

\author{Long Liu}
\affiliation{Institute of Microelectronics, Chinese Academy of Sciences, Beijing, 100029, China}   
\affiliation{University of Chinese Academy of Sciences, Beijing, 100049, China}

\author{Guozhong Xing}
\email{gzxing@ime.ac.cn} 
\affiliation{Institute of Microelectronics, Chinese Academy of Sciences, Beijing, 100029, China}    
\affiliation{University of Chinese Academy of Sciences, Beijing, 100049, China}

\author{Xing-Qiu Chen}
\email{xingqiu.chen@imr.ac.cn}
\affiliation{Shenyang National Laboratory for Materials Science, Institute of Metal Research,Chinese Academy of Sciences, Shenyang 110016, China.}
\affiliation{School of Materials Science and Engineering, University of Science and Technology of China, Shenyang 110016, China.}

\author{Yan Sun}
\email{sunyan@imr.ac.cn}
\affiliation{Shenyang National Laboratory for Materials Science, Institute of Metal Research,Chinese Academy of Sciences, Shenyang 110016, China.}
\affiliation{School of Materials Science and Engineering, University of Science and Technology of China, Shenyang 110016, China.}

\keywords{Altermagnetic material, Spin current, First principle calculation}

\begin{abstract} 

%\remove{
%Altermagnets represent an emerging class of materials in condensed matter physics, combining an 
%antiferromagnetic ground state with spin-splitting effects typically found in ferromagnets. This 
%unique duality enables ultrafast spin-dependent responses, offering new opportunities for spin-current 
%generation—a critical advancement for overcoming the limitations of conventional spin-transfer and 
%spin-orbit torques in magnetic memory technologies.}
Altermagnets combine antiferromagnetic order with ferromagnet-like spin splitting, a duality that unlocks ultrafast spin-dependent responses. This unique property creates unprecedented opportunities for spin-current generation, overcoming the intrinsic limitations of conventional spin-transfer and spin-orbit torque approaches in magnetic memory technologies.
%\remove{In this work}
Here, we establish a fundamental 
relationship between Fermi surface geometry and time-reversal-odd ($\mathcal{T}$-odd) spin currents in altermagnets through combined model analysis and first-principles calculations.
We demonstrate that a $d$-wave altermagnet with a flat Fermi surface can achieve a theoretical upper limit of 
charge-to-spin conversion efficiency (CSE) of 100\%. This mechanism is realized in the newly discovered 
room-temperature altermagnetic metal KV$_2$O$_2$Se, which exhibits a CSE
of $\sim$78\% at the charge neutrality point—nearly double that of RuO$_2$, setting a new record for 
$\mathcal{T}$-odd CSE. Under electron doping, this efficiency further increases to $\sim$98\%, 
approaching the theoretical limit. Our work advances the fundamental understanding of $\mathcal{T}$-odd spin currents 
via Fermi surface geometry engineering and provides key insights for developing next-generation 
altermagnet-based memory devices.

\end{abstract}

\maketitle
%\section{Introduction}
\textit{Introduction}---Fermi surface geometry plays a fundamental role in determining the transport properties 
of materials and driving quantum phase transitions \cite{simon2013oxford}. The interplay 
between Fermi surface geometry, topology, and strong electronic correlations has led to 
many intriguing phenomena in condensed matter physics and material science \cite{weng2015,si2001}. 
Specifically, Fermi surface nesting can induce electronic instabilities, giving rise to emergent 
quantum states such as charge density waves, spin density waves, and unconventional 
superconductivity \cite{tan2021,arachchige2022,fawcett1988,murayama1997,bardeen1957,terashima2009}. 
In Dirac and Weyl semimetals, the point-like Fermi surfaces are responsible for ultrahigh carrier 
mobility, strong anomalous Hall effect, and giant magnetoresistance \cite{shekhar2015,vistoli2019,
novak2015}. Furthermore, the linear dispersion crossing the Fermi surface in type-II Weyl semimetals 
breaks Lorentz symmetry, resulting in a chiral anomaly transport response that exhibits strong 
directional dependence relative to the applied electric field \cite{xiao2010,hasan2010,zhang2018d}. 
Materials with open, concave, single-band Fermi surface can exhibit goniopolar behavior, manifesting 
as opposite conduction polarities along in-plane and cross-plane directions, as observed in the 
Seebeck and Hall coefficient measurements \cite{he2019a,pan2022}.

Beyond these exotic electrical responses and quantum phase transitions, the interplay  
between Fermi surface geometry and magnetic order plays a key role in spin current generation—a 
central challenge in modern spintronics \cite{freimuth2014,zzelezny2017,naka2019,hayami2020,
yuan2020,naka2021,gonzalez-hernandez2021,ma2021,bose2022,bai2022,bai2023,cui2023a,ssmejkal2022,shao2023,
hu2025cataloguecpairedspinmomentumlocking}. 
Altermagnets, a newly discovered class of magnetic materials at the frontier of spintronics, combine an 
antiferromagnetic ground state with momentum-dependent spin splitting similar to that of ferromagnets and 
noncollinear antiferromagnets (AFM) \cite{ssmejkal2022a,bai2024,song2025,guo2023b,liu2024a,liu2024b,
kriegner2016,kriegner2017}. This unique duality positions altermagnets as an ideal platform for 
AFM-based spintronics, particularly for generating time-reversal-odd ($\mathcal{T}$-odd) longitudinal spin-polarized 
currents, a key functionality for next-generation spintronic memory devices \cite{freimuth2014,zzelezny2017,naka2019,
hayami2020,yuan2020,naka2021,gonzalez-hernandez2021,ma2021,bose2022,bai2022,bai2023,cui2023a}. 
Owing to their negligible net magnetization, altermagnets exhibit no stray fields and are robust 
against external magnetic perturbations, enabling high-density integration of spintronic devices
\cite{ssmejkal2022a,bai2024,song2025}. Moreover, because $\mathcal{T}$-odd longitudinal spin currents originate 
from spin splitting in momentum space, they are expected to possess significantly longer spin diffusion 
lengths compared to spin–orbit coupling (SOC)-induced time-reversal-even spin Hall effect \cite{zzelezny2017}.

In this work, we establish a fundamental connection between Fermi surface geometry and the charge-to-spin 
conversion efficiency (CSE) in altermagnets through effective model Hamiltonian analysis and first-principles 
calculations. We demonstrate that the non-relativistic charge-to-spin conversion is governed by the spin 
anisotropy of the Fermi surface. Remarkably, in the extreme case of spin-splitting anisotropy, $d$-wave 
altermagnet with flat Fermi surfaces throughout the entire Brillouin Zone (BZ), the CSE for $\mathcal{T}$-odd 
spin currents can reach the theoretical upper limit of 100\%. Guided by this principle, 
we identify the recently discovered altermagnet KV$_2$Se$_2$O \cite{jiang2025} as a near-ideal candidate, 
with a Fermi surface closely resembling this extreme case of spin-splitting anisotropy. First-principles 
calculations reveal that KV$_2$Se$_2$O achieves a CSE nearly double that of RuO$_2$
\cite{gonzalez-hernandez2021} at the charge-neutral point, setting a new record for the $\mathcal{T}$-odd CSE. 
Moreover, CSE can reach approximately 98\% in the electron 
doping zone, approaching the theoretical limit. Our findings reveal that Fermi surface geometry as a 
fundamental design parameter for optimizing $\mathcal{T}$-odd spin currents, offering new insights for the development 
of altermagnet-based memory technologies.

\begin{figure}[htb]
	\centering
	\includegraphics[scale=0.42]{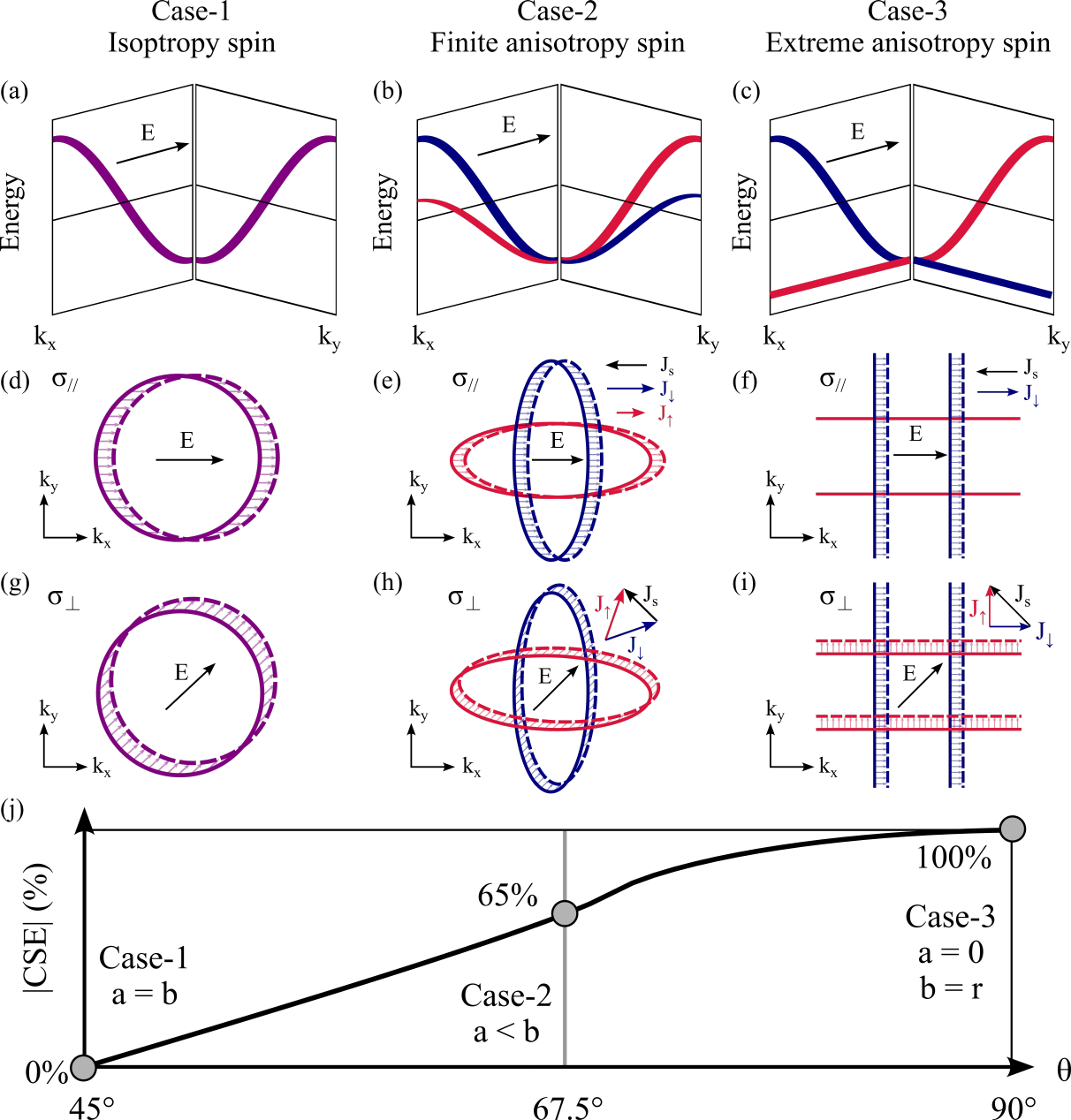}
	\caption{Schematic illustration of altermagnets with different 
		spin-splitting anisotropy based on a model study. 
		(a-c) Band structures and (d-i) corresponding Fermi surfaces of 
		altermagnet with different degrees of spin-splitting anisotropy.
		Longitudinal conductivity $\sigma_{\vert\vert}$ is generated in (d-f) with $E$//[100] while 
		transverse conductivity $\sigma_{\perp}$ is generated in (g-i) with $E$//[110]. 
		The spin currents from spin-up ($J_\uparrow$) and spin-down ($J_\downarrow$) are marked by
		red and blue, respectively. The overall spin current is given by $J_s=J_\uparrow-J_\downarrow$.
		(j) The calculated $\lvert$CSE$\rvert$ for different polar angles $\theta$, which control
		the spin-splitting anisotropy in a minimal altermagnet model.
		Case-1 ($a=b$) corresponds to an antiferromagnetic system with isotropic spin orientations.
		Case-2 ($a<b$) corresponds to a general case of altermagnetic system with non-equivalent
		spin orientations along $k_x$ and $k_y$.
		Case-3 ($a=0, b=r$) is an extreme case with spin polarization oriented purely along a single direction.
		Model parameters of $\varepsilon_0 = 1.2~\text{eV}$ and $r = 2~\text{eV}$ are used to simulate 
		varying degrees of spin-splitting anisotropy.
		} \label{fig1}
\end{figure}%

%\section{Results and discussion}
\textit{Results and discussion}---To capture the momentum-dependent anisotropic spin splitting, 
we first reformulated the $d$-wave altermagnetic 
model Hamiltonian \cite{roig2024,yuan2020,hayami2020}, which reads
\begin{equation}\label{eq:H}
	\begin{aligned}
	H(\mathbf{k}) = \varepsilon_{0}-\begin{pmatrix}
		acosk_x+bcosk_y & 0 \\
		0 & bcosk_x+acosk_y
		\end{pmatrix},
	\end{aligned}
\end{equation}
where the SOC is not included due to its negligible effect on collinear spin-splitting anisotropy in altermagnets. 
By tuning the model coefficients $a$ and $b$, the anisotropy between bands with different spin orientations can 
be modulated. Specifically, when $a = b$, the model describes a collinear antiferromagnetic system in which spin-up 
and spin-down electrons are fully degenerate throughout the BZ [Fig.~\ref{fig1}(a,d)]. As $a$ and $b$ become different, 
a spin splitting $\Delta = (a - b)(\cos k_x - \cos k_y)$ emerges between bands with different spin orientations. This leads to anisotropic manifestations of the Fermi surfaces in momentum space [Fig.~\ref{fig1}(b,e)]. Notably, when either 
$a$ or $b$ becomes zero, the system exhibits maximal spin-splitting anisotropy. In this extreme configuration, only one 
spin orientation exhibits a significant Fermi velocity along a specific direction (e.g., $k_x$), while the other has nearly zero Fermi velocity [Fig.~\ref{fig1}(c)]. Consequently, the Fermi surfaces no longer appear 
as closed bubble in momentum space, but present as several mutually perpendicular two-dimensional (2D) straight lines 
or three-dimensional (3D) flat surfaces [Fig.~\ref{fig1}(f)]. Such $d$-wave flat Fermi surfaces represent complete spin-channel 
separation in altermagnet. 

\begin{figure*}[htb]
	\centering
	\includegraphics[scale=0.9]{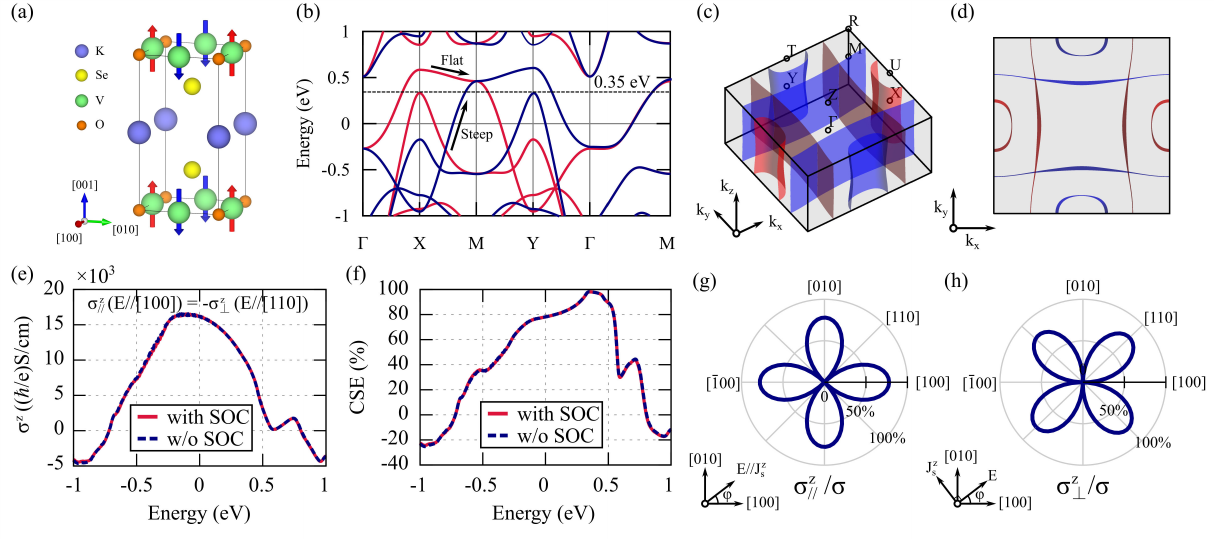}
	\caption{Crystal structure, band structure, and spin current characteristics of KV$_2$Se$_2$O.
		(a) Crystal structure and spin configuration of KV$_2$Se$_2$O.
		(b) Spin-resolved band structure without spin–orbit coupling (SOC); red and blue lines indicate different spin orientations.
		(c, d) Fermi surface of KV$_2$Se$_2$O (c) 3D view, (d) top view. Red and blue surfaces denote opposite spin orientations.
		(e, f) Longitudinal ($\sigma_{\vert\vert}^z/\sigma$) and transverse ($-\sigma_{\perp}^z/\sigma$) CSE values under electric fields applied along [100] and [110] directions, respectively, as functions of chemical potential.
		(g, h) Angular dependence of the longitudinal and transverse spin current conversion efficiency at the charge neutrality point with $\eta = 10$~meV, showing periodic variation as the electric field rotates within the $xy$ plane.
	}
	\label{fig2}
\end{figure*}

Under an applied electric field $E$, the non-equilibrium redistribution of the Fermi surfaces with different spin 
orientations lead to $\mathcal{T}$-odd spin-polarized current. Within the single-particle approximation, the corresponding 
spin conductivity tensor is given by \cite{gonzalez-hernandez2021,freimuth2014}:
\begin{equation}\label{eq:sc}
	\begin{aligned}
\sigma_{ij}^{s_k} =& - \frac{e\hbar}{V\pi N_\mathbf{k}}\sum_{\mathbf{k}} \tilde{\sigma}_{ij}^{s_k}(\mathbf{k}),\\
	\end{aligned}
\end{equation}
\begin{equation}\label{eq:local}
	\begin{aligned}
		\tilde{\sigma}_{ij}^{s_k}(\mathbf{k}) =& \sum_{mn}
\frac{   \text{Re} (
	\braket{u_{n\mathbf{k}}|\hat{J}^{s_k}_{i}|u_{n\mathbf{k}}}
	\braket{u_{n\mathbf{k}}|\hat{v}^j_{n}|u_{m\mathbf{k}}})\eta^2
}
{
	((E_F-E_{n\mathbf{k}})^2+\eta^2)((E_F-E_{m\mathbf{k}})^2+\eta^2)
},
\end{aligned}
\end{equation}
where
$\hat{J}^{s_k}_i = \frac{1}{2} \left\lbrace \hat{s}_k,\hat{v}_i \right\rbrace $ is the spin operator with spin polarization 
along $k$ and current flow along $i$. $j$ denotes the direction of the applied field. $\eta$ is a broadening parameter associated 
with the relaxation time. A dense $\mathbf{k}$-grid of 320$^3$ is employed for the integration over BZ and has been tested to be fully converged 
in the subsequent discussion. To quantify spin current generation under varying spin-splitting anisotropy, we parameterize the 
Hamiltonian in Eq.~\eqref{eq:H} to compute the spin conductivity. For convenience, the coefficients $a$ and $b$ are parameterized in terms 
of polar coordinates through $a=r\cos\theta$ and $b = r\sin\theta$. By gradually varying the angle $\theta$ from  \SI{45}{\degree} to \SI{90}{\degree}, a 
continuous tuning of the spin-splitting anisotropy in the altermagnets can be achieved. 
The CSE of the magnetic materials can be quantified as $\text{CSE} = \sigma^{s_k}_{ij}/\sigma_{ii} \times 100\%$.
 The CSE of different spin-splitting anisotropies is plotted by tuning the polar 
angle $\theta$ of the model [Fig.~\ref{fig1}(j)].

At $\theta=\SI{45}{\degree}$, the spin splitting is isotropic, and the CSE in such 
non-relativistic system is forbidden. As the angle $\theta$ changes from \SI{45}{\degree} to \SI{90}{\degree}, the CSE increases monotonically, corresponding to the Case-2 
in Fig.~\ref{fig1}. Remarkably, a theoretical maximum CSE of 100\% can be achieved in Case-3 of Fig.~\ref{fig1}, with an extreme 
spin-splitting anisotropy at $\theta = \SI{90}{\degree}$.  At this anisotropic limit, the Fermi surfaces of electrons with 
different spin orientations become mutually perpendicular straight lines in 2D, or flat surfaces in 3D. These results 
establish Fermi surface geometry and spin-splitting anisotropy as the primary determinants of spin-current conversion 
efficiency in altermagnets.

Guided by the insights from the effective model analysis, we now examine KV$_2$Se$_2$O, a recently synthesized 
room-temperature altermagnetic metal \cite{jiang2025}, as a realistic platform to approach the theoretical limit 
of CSE. Spin-resolved ARPES and neutron diffraction measurements confirm that KV$_2$Se$_2$O hosts 
$d$-wave altermagnetic order, with the magnetic order localized on V atoms and the Néel vector aligned along the $c$-axis 
[Fig.~\ref{fig2}(a)]. First-principles calculations performed in the FPLO code \cite{koepernik1999,koepernik2023,opahle1999,
perdew1996} reproduce the electronic band structure and Fermi surface, in excellent 
agreement with experimental ARPES data \cite{jiang2025}.

Note that a significant anisotropic feature proposed as Case-3 in Fig.~\ref{fig1}(c) can be found along the $X$-$M$-$Y$ path. 
Taking the $X$-$M$ path as an example, the spin-up channel is almost flat while the spin-down channel exhibits a 
significant Fermi velocity, [see the black arrows in Fig.~\ref{fig2}(b)]. The Fermi surface of KV$_2$Se$_2$O exhibits 
quasi-two-dimensional characteristics, with its shape remaining consistent across different $k_z$-planes 
[Fig.~\ref{fig2}(c-d)]. Moreover, the Fermi surface consists of two groups of electrons with different spin 
orientations. One group appears as two elliptical cylinders extending throughout the entire $k_z$-plane near 
the boundaries of the BZ along either $k_x$ or $k_y$ direction. This ellipticity results in a finite spin-splitting,
corresponding to Case-2 in Fig.~\ref{fig1}(e). The other group forms two flat planes perpendicular 
to either the $k_x$ or $k_y$ direction, which is very similar to Case-3 in Fig.~\ref{fig1}(f) with 
an extreme spin-splitting anisotropy. 

According to the effective model analysis, when the Fermi surface of the system satisfies the Case-3 in 
Fig.~\ref{fig1}(c), a theoretical upper limit of the CSE 
can be reached. Therefore, we further 
investigate the CSE in KV$_2$Se$_2$O. 

Since the SOC effect of KV$_2$Se$_2$O is relatively small near the Fermi level, the $\sigma^x_{ij}$ 
and $\sigma^y_{ij}$ components of the spin current are negligible compared to the $\sigma^z_{ij}$.
Besides, symmetry constraint only allows $\sigma^z_{ij}$ induced by an in-plane electric field.
In this case, the spin current tensors under a rotating in-plane electric field are further evaluated 
by a series of matrix rotations \cite{seemann2015}, which reads
 \begin{equation}
 	\begin{aligned}
 		\sigma_{i'j'}^{s_{k'}}=\sum_{ii'}\sum_{jj'}\sum_{kk'}D_{ii'}D_{jj'}D_{kk'}\sigma_{ij}^{s_k},
 	\end{aligned}
 \end{equation}
where $D_{ii'}$ is the Euler rotation matrix defined by the in-plane rotation angle ($\varphi$) of 
the applied field, $i'$ denotes the direction aligned with the rotated electric field.
Both transverse and longitudinal spin currents can be induced when the electric field rotates within the $xy$ plane.
To distinguish them, we write them as $\sigma_{\perp}^z(E\vert\lvert x')=\sigma_{y'x'}^z(E\vert\lvert x')$ and $\sigma_{\vert\vert}^z=\sigma_{x'x'}^z(E\vert\vert x')$ 
in the following
discussion. The maximum value of the transverse and longitudinal spin current can be achieved when the electric 
field is either parallel to the [110] or [100] direction, satisfying the relation
$\sigma_{\perp}^z(E\vert\vert[110])=-\sigma_{\vert\vert}^z(E\vert\vert[100])$.
For convenience, we use $\sigma_{\perp}^z$ and $\sigma_{\vert\vert}^z$ to refer to these peak values.

As shown in Fig.~\ref{fig2}(c), 
the magnitude of transverse ($\sigma_{\perp}^z$) and longitudinal ($\sigma_{\vert\vert}^z$)
spin currents exceed 1.6$\times$10$^4$($\hbar$/e)~S/cm at the charge neutrality point with $\eta=10$~meV, 
which is much higher than the reported result in RuO$_2$\cite{gonzalez-hernandez2021} (around 
5$\times$10$^3$($\hbar$/e)~S/cm with $\eta=10$~meV). This large value arises from the 
high Fermi velocity induced by strong spin splitting. Furthermore, due to the substantial spin 
splitting in KV$_2$Se$_2$O, this spin current can exceed 1.0$\times$10$^4$($\hbar/e$)~S/cm 
over a broad chemical potential range (\SIrange{-0.4}{0.5}{eV}). This could significantly 
facilitate the experimental detection.

Based on the spin current results, we also calculated the CSE =
$\sigma_{\perp}^z/\sigma\times100~\%$ at
different chemical potentials [Fig.~\ref{fig2}(e)]. One can see that the CSE of KV$_2$Se$_2$O reaches 
\SI{78}{\percent} at the charge neutrality point and goes up to \SI{98}{\percent} when the chemical potential 
approaches 0.35~eV, nearly approaching the theoretical upper limit. The remaining incomplete conversion can be 
attributed to the contributions from the cylindrical Fermi surface near the BZ boundary.  
As the Fermi level shifts to 0.35~eV, this pair of spin-polarized Fermi surfaces gradually vanishes
[see the dashed line in Fig.~\ref{fig2}(b)]. At this point, the Fermi surface is almost entirely determined by two 
mutually perpendicular flat planar surfaces. Based on the results in Fig.~\ref{fig1}, such flat Fermi surface 
configuration in $d$-wave altermagnet allows for a CSE approaching \SI{100}{\percent}.

In Fig.~\ref{fig2}(g-h), the spin current and corresponding CSE are simulated under a 
rotating electric field.
The CSEs of $\sigma_{\vert\vert}^{z}$ and $\sigma_{\perp}^{z}$ follow cosine and sine functional forms
with period $\pi$ with respect to the rotation angle $\varphi$, respectively [Fig.~\ref{fig2}(g,h)].
This implies that the maximum CSE can be switched from longitudinal to transverse direction
as the electric fields rotates from [100] to [110] direction.

\begin{figure}[htb]
	\centering
	\includegraphics[scale=0.6]{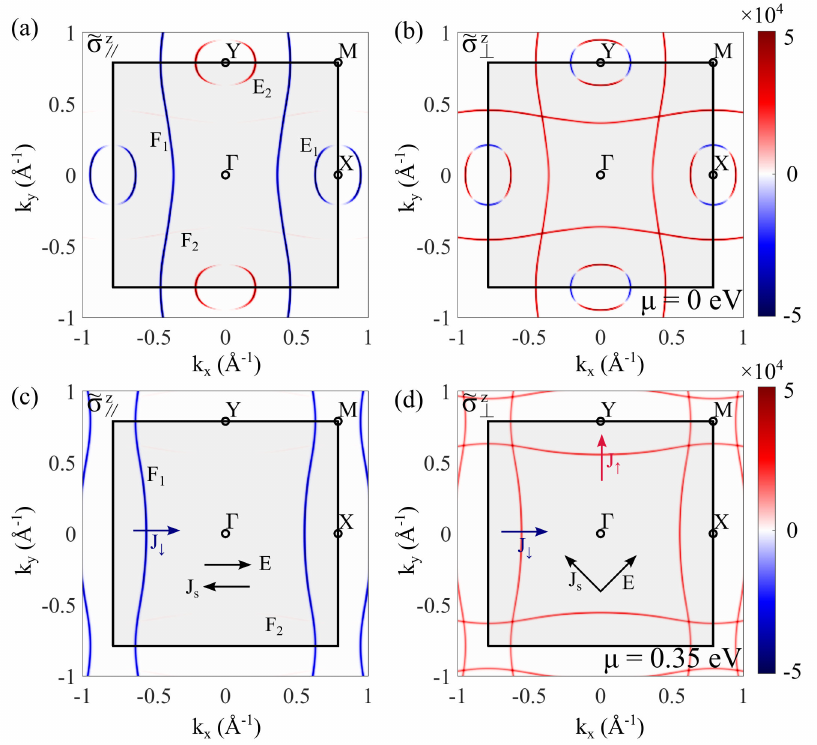}
	\caption{Spin current distributions for different conductivity components.
		(a) $\tilde{\sigma}_{\vert\vert}^z$ and (b) $\tilde{\sigma}_{\perp}^z$ at the charge neutrality point.
		(c) $\tilde{\sigma}_{\vert\vert}^z$ and (d) $\tilde{\sigma}_{\perp}^z$ at a chemical potential $\mu = 0.35$~eV.
		Color bars are in the unit of ($\hbar$/e)\si{\angstrom^2}. The black square denotes the BZ.
	}
	\label{fig3}
\end{figure}

To further understand the origin of this large CSE, we calculated the momentum-space distribution of the spin current 
[Fig.~\ref{fig3}]. Since the 
contributions from different $k_z$ planes are nearly identical, we use the $k_z=0$ plane as a representative 
case. As shown in Fig.~\ref{fig3}(a), both $\sigma_{\vert\vert}^z$ and $\sigma_{\perp}^z$ 
primarily consist of 
four parts: two elliptical cylinders centered at the $X$ and $Y$ points (marked as E$_1$ and E$_2$), 
as well as two sets of nearly flat surfaces perpendicular to the $k_x$ and $k_y$ directions
(marked as F$_1$ and F$_2$), corresponding to the Fermi surfaces in Fig.~\ref{fig2}(c-d).

Here we take the longitudinal current as an example, as the transverse
part can be understood through a similar method with spin flow along [1$\bar{1}$0] direction.
When electric field is aligned with [100] [Fig.~\ref{fig3}(a,c)], the induced polarization from spin-up 
electrons in the elliptical cylinder E$_1$ is generated along the long axis while the 
polarization induced from spin-down electrons in E$_2$ is generated along the short axis.
The combination of these two polarizations with different spin orientations results in a 
cancellation effect for the generation of spin current. Since the conductivity $\sigma$ 
receives contributions with the same sign from both E$_1$ and E$_2$, this cancellation 
effect suppresses the overall CSE. Regarding the flat Fermi surface, since F$_1$ and F$_2$ 
are oriented either perpendicular or parallel to the $k_x$ direction, the spin current 
is predominantly driven by the polarization of spin-up electrons in F$_1$, while the 
polarization of spin-down electrons in F$_2$ hardly generates a opposing current. 
This would drive the overall CSE to the theoretical limit.

As the Fermi level shifts to the $\sim$0.35 eV, it approaches the
extreme case, where the Fermi surfaces E$_1$ and E$_2$ gradually vanish
and only the flat Fermi surfaces F$_1$ and F$_2$ persist [Fig.~\ref{fig3}(c-d) and 
Fig.~\ref{fig2}(b)].
In this case, the origin of the spin current 
is nearly identical to that of electrical conductivity, resulting in an almost complete CSE
as presented in Fig.~\ref{fig2}(f).
\begin{figure}[htb]
	\centering
	\includegraphics[scale=0.5]{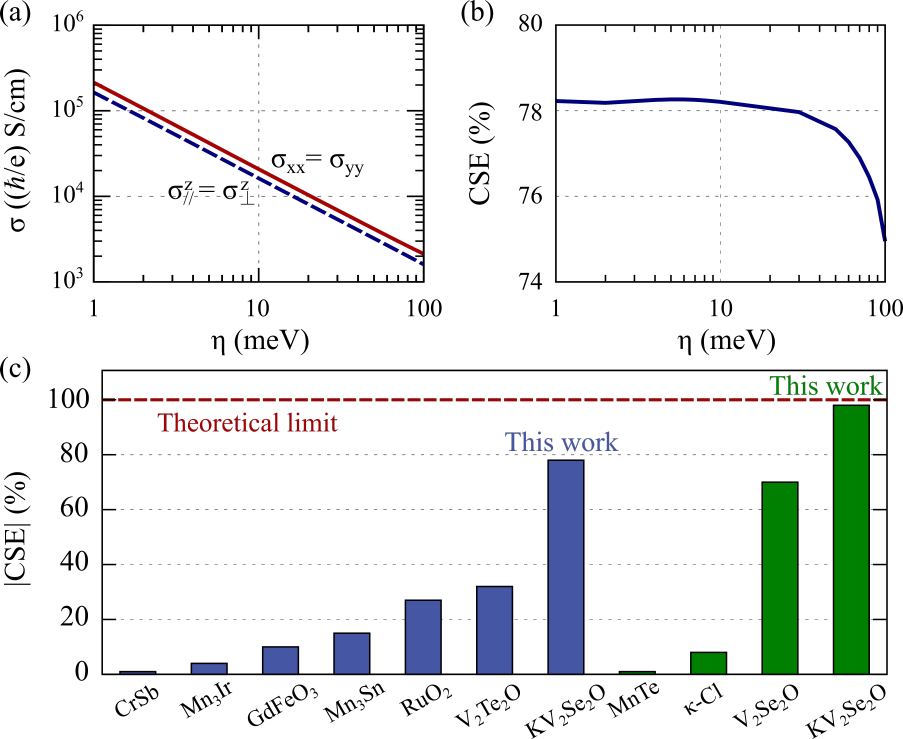}
	\caption{Conductivity, spin conductivity, and CSE of KV$_2$Se$_2$O, in comparison with other altermagnets. 
		(a) $\sigma_{xx}$, $\sigma_{yy}$, $\sigma_{\vert\vert}^{z}$, and $\sigma_{\perp}^{z}$ of KV$_2$Se$_2$O under different $\eta$ values.
		(b) CSE of KV$_2$Se$_2$O at various $\eta$ values.
		(c) Absolute CSE of KV$_2$Se$_2$O compared with other altermagnets at the charge neutrality point (blue bars) and at a finite doping level (green bars). Data sources: CrSb (this work, 
		calculated using identical methodology) \cite{zhou2025,reimers2024,ding2024,li2024,yu2025}, Mn$_3$Ir 
		and Mn$_3$Sn \cite{zzelezny2017}, GdFeO$_3$ \cite{naka2021}, RuO$_2$ and MnTe \cite{gonzalez-hernandez2021}, 
		V$_2$Te$_2$O \cite{cui2023a}, $\kappa$-Cl \cite{naka2019}, and V$_2$Se$_2$O \cite{ma2021}.
	}
	\label{fig4}
\end{figure}

The conductivity and the spin conductivity are further evaluated with a 
varying $\eta$ since Eq.~\eqref{eq:sc} may vary under different relaxation times [Fig.~\ref{fig4}(a)]. 
Both the conductivity $\sigma$ and the spin conductivity 
$\sigma_{\vert\vert}^{z}$ and $\sigma_{\perp}^{z}$ are inversely 
proportional to $\eta$ within a reasonable range between 
1~meV and 100~meV. Correspondingly, the CSE in KV$_2$Se$_2$O remains nearly unchanged within this relaxation times, 
consistently within the range of \SIrange{75}{78}{\percent} [Fig.~\ref{fig4}(b)]. This indicates that 
the CSE is highly robust against variations in $\eta$.
Besides, since the connection between CSE and spin current originates from the intrinsic Fermi surface geometry rather than the details of scattering, the predictions from both model analysis and first-principles calculations are quantitatively reliable.

A comparison of CSE values at the charge neutrality point highlights KV$_2$Se$_2$O’s superiority [Fig.~\ref{fig4}(c)]. 
Its $d$-wave flat Fermi surface and extreme spin-splitting anisotropy enable a CSE nearly twice that of other 
altermagnets. Previous studies suggest chemical doping or thermal gradients could further enhance spin current 
conversion rates approaching 70\% in an insulator like monolayer V$_2$Se$_2$O \cite{ma2021}. We believe that these 
approaches can also be applied to KV$_2$Se$_2$O to achieve an almost complete spin current conversion. Compared 
to previous reports, KV$_2$Se$_2$O with stronger $\mathcal{T}$-odd current and higher CSE, 
emerges as an ideal material platform for high-efficiency spintronic devices. 

%\section{Conclusion}
%\remove{
%\textit{Conclusion}---In summary, we have established a fundamental connection 
%between Fermi surface geometry and $\mathcal{T}$-odd spin currents in altermagnets. 
%Through effective model analysis, we demonstrate that when the spin-splitting anisotropy of an 
%altermagnet reaches its limit, the Fermi surfaces for different spin orientations become mutually 
%perpendicular straight lines or flat planes. This enables a theoretical upper limit of spin current 
%conversion rate of 100\%. 
%First-principles calculations reveal that KV$_2$Se$_2$O can generate a spin 
%current exceeding 1.6$\times$10$^4$($\hbar$/e)~S/cm under $\eta=10$~meV and maintain a 
%robust spin current conversion ratio above 
%75\% under reasonable relaxation times, establishing a new record.
%Moreover, the CSE of KV$_2$Se$_2$O could even reach 98\% after electron doping, 
%approaching the theoretical upper limit. 
%This work provides fundamental insights 
%into $\mathcal{T}$-odd spin currents by establishing their intrinsic relationship 
%with Fermi surface geometry, while demonstrating KV$_2$Se$_2$O as an exceptional altermagnetic 
%material that achieves 
%near-unity CSE—a critical advancement for developing next-generation 
%altermagnet-based spintronic memory devices.
%}
%\add{
\textit{Conclusion}---In conclusion, we have established a universal route to maximize charge-to-spin conversion efficiency by leveraging the flat Fermi surface in a $d$-wave altermagnet.
This fundamental connection between $\mathcal{T}$-odd spin currents and Fermi surface geometry in altermagnets 
has been unequivocally unveiled through effective model analysis, 
demonstrating the significant role of the flat Fermi surfaces associated with the spin-splitting anisotropy of the altermagnet 
in achieving the theoretical limit of CSE.
Our model prediction has been computationally realized in a room-temperature altermagnetic metal KV$_2$Se$_2$O with characteristic flat Fermi surfaces.
First-principles calculations reveal that KV$_2$Se$_2$O can generate a spin current exceeding 1.6$\times$10$^4$($\hbar$/e)~S/cm under $\eta=10$~meV 
and maintain a robust CSE above 75\% under reasonable relaxation times. 
This CSE value is nearly twice that of RuO$_2$ and represents the highest value reported among altermagnets to date.
Moreover, upon appropriate electron doping, the CSE value can be further enhanced,
approaching a near-theoretical limit of 98\% when the chemical potential reaches 0.35 eV.
This record-breaking CSE is a significant milestone, offering a promising avenue for the development of next-generation high-performance altermagnet-based spintronic memory devices.
We emphasize that the relationship between $\mathcal{T}$-odd spin currents and Fermi surface geometry identified in this work 
is general, providing a valuable design principle for high-throughput discovery of new materials with enhanced charge-to-spin conversion capabilities.
%}

\begin{acknowledgments}
This work was supported by the Liao Ning Revitalization Talents Program (Grant No. XLYC2203080), 
Natural Science Foundation of China (Grants No. 52271016, 52188101, 52422112, 62074164 and 62374180), 
National Key Research and Development Program of China (Grant No. 2021YFB3501503, 2021YFB3601300), 
the National Chinese Academy of Sciences Project for Young Scientists in Basic Research, Grant 
No. YSBR-109, the Special Projects of the Central Government in Guidance of Local Science and 
Technology Development(2024010859-JH6/1006). 
\end{acknowledgments}

\appendix


%\bibliography{alterm.bib}

\end{document}